\documentclass{svproc}
\usepackage{url}
\usepackage[pdftex]{graphicx}

\usepackage[labelfont=bf]{caption}

\usepackage{float}				
\usepackage{wrapfig} 		
\usepackage{subfig}			
\usepackage{xspace}
\usepackage{booktabs} 	
\usepackage{arydshln}		
\usepackage{amsfonts}

\newcommand{\numnodes}{\ensuremath{7,178}\xspace}
\newcommand{\numedges}{\ensuremath{25,104}\xspace}
\newcommand{\COREsize}{\ensuremath{297}\xspace}
\newcommand{\muSPLwww}{\ensuremath{4.27}\xspace}
\newcommand{\muSPLdw}{\ensuremath{4.35}\xspace}
\newcommand{\muSPLcore}{\ensuremath{3.97}\xspace}
\newcommand{\percentzeroOD}{\ensuremath{87\%}\xspace}

\begin{document}
\mainmatter              
\title{The darkweb: a social network anomaly}
\titlerunning{Darkweb}  
%

\author{Kevin P. O'Keeffe \inst{1} \and Virgil Griffith\inst{2}
Yang Xu \and Paolo Santi \inst{1} \and Carlo Ratti \inst{1} }
\authorrunning{Kevin P. O'Keeffe et al.} 
%
\tocauthor{Ivar Ekeland, Roger Temam, Jeffrey Dean, David Grove,
Craig Chambers, Kim B. Bruce, and Elisa Bertino}
\institute{SENSEable City Laboratory \\ Massachusetts Institute of Technology \\ Cambridge, MA 02139, USA\\
\and
Ethereum Foundation
\and
Hong Kong Polytechnic University
}

\maketitle              

\begin{abstract}
We analyse the darkweb and find its structure is unusual. For example, $ \sim 87 \%$ of darkweb sites \emph{never} link to another site. To call the darkweb a ``web'' is thus a misnomer -- it's better described as a set of largely isolated dark silos. As we show through a detailed comparison to the World Wide Web (www), this siloed structure is highly dissimilar to other social networks and indicates the social behavior of darkweb users is much different to that of www users. We show a generalized preferential attachment model can partially explain the strange topology of the darkweb, but an understanding of the anomalous behavior of its users remains out of reach. Our results are relevant to network scientists, social scientists, and other researchers interested in the social interactions of large numbers of agents.



\keywords{social networks, graph theory, computational network science}
\end{abstract}

\section{Introduction}

Studies of the World Wide Web (www) have had much impact. An understanding of its topology have let us better understand how its information is stored and can be retrieved \cite{albert1999diameter,barabasi2000scale}. Insight into its paradoxical resilience \cite{albert2000error} have allowed us to design more fault tolerant networks \cite{wang2007fault}, and the universal dynamics of it growth patterns \cite{barabasi1999emergence} have shed light on the behaviors of many others classes of networks \cite{barabasi2003scale}. And perhaps most importantly, being one of the earliest studied networks\footnote{alongside smallworld networks \cite{watts1998collective}.} \cite{albert1999diameter,barabasi2000scale}, the www helped give rise to the flame of attention that network science enjoys today. 

This paper is about the www's shady twin: the darkweb. Though the darkweb is similar to the www, both being navigated through a web browser, a key difference the two is that the identities of darkweb users are hidden -- that's what makes it `dark'. This gives the darkweb an infamous air of mystery, which, along with the sparse academic attention it has received \cite{everton2012disrupting,darknet}, makes it ripe for analysis. And beyond satisfying curiosity, its reasonable to think studies of the darkweb could have as much applied impact as the studies of the www. Insight about the structure or dynamics of the darkweb, for instance, could potentially allow policy-makers to better police its more sinister aspects. 

There are many natural questions to ask about the darkweb. Does it have the same topology as the www, and therefore hold and distribute information in the same way? Is it resilient to attack? Social questions can also be posed. Results from psychology and game theory show people behave more socially when they are watched \cite{bradley2018does,kristofferson2014nature} or have reputations to uphold \cite{nowak2011supercooperators}. Do darkweb users, with their masked faces, therefore behave more selfishly than www users, whose faces are bare? 

Here we address some of these questions by performing a graph theoretical analysis of the darkweb (we define exactly what we mean by the darkweb below) and comparing it to the www. We find the topologies of the darkweb and the www are starkly different -- in fact, the darkweb is much different to many social networks -- and conjecture this is due to their users' differing social behaviour. We hope our work stimulates further research on the darkweb's structure as well as the social dynamics of its users.

\section{Data Collection}
There is no single definition of the darkweb. It is sometimes loosely defined as ``anything seedy on the Internet'', but in this work we define the darkweb as all domains underneath the ``.onion'' psuedo-top-level-domain\cite{rfc7686} (which is sometimes called the onionweb). Specifically, we mean the subset of the web where websites are identified not by a human-readable hostname (e.g., yahoo.com) or by a IP number (e.g., 206.190.36.45), but by a randomly generated 16-character address (specifically, a hash fingerprint).  Each website can be accessed via its hash, but it is very difficult to learn the IP number of a website from its hash -- this is what makes the web `dark';  without an IP number, it is exceedingly difficult to trace the geographical origin of a communication.

Crawling the darkweb is not much harder than crawling the regular web.  In our case, we crawled the darkweb through the popular tor2web proxy onion.link.  Onion.link does all of the interfacing with Tor, and one can crawl all darkweb pages without a login simply by setting a standard crawler to specifically crawl the domain \texttt{*.onion.link}. Darkweb pages are written in the same HTML language as the regular web which means we could crawl onion.link using the commercial service scrapinghub.com. Starting from two popular lists of darkweb sites,\footnote{\texttt{http://directoryvi6plzm.onion} and \texttt{https://ahmia.fi/onions/}} we accessed each page and crawled all linked pages using breadth-first search.

Most analyses of the www are at the page-level, where each node is an individual URL. One could adopt this model for the darkweb. It would be a natural choice for engineering motivated research question, such as studying crawlability. But the page-level model is not natural for socially motivated research questions, which we are interested in in this study, because the page-level graph is influenced more by the various choices of content management system, rather than by social dynamics. So we instead follow the modeling choice in \cite{lehmberg2014graph} and \emph{aggregate by second-level domain} (for the onionweb the second-level domain is equivalent to \cite{lehmberg2014graph}'s ``pay-level domain'').  This means that links within a second-level domain are ignored as socially irrelevant self-connections.  In this formulation, each vertex in the darkweb is a \emph{domain} and every directed edge from $u \rightarrow v$ means there exists a page within domain $u$ linking to a page within domain $v$.  The weight of the edge from $u \rightarrow v$ is the number of pages on domain $u$ linking to pages on domain $v$.

The Tor Project Inc. (the creators and custodians of the darkweb) state there are $\sim \!\! 60,000$ distinct, active .onion addresses \cite{tormetrics}. In our analysis, however, we found merely \numnodes active .onion domains.  We attribute this high-discrepancy to various messaging services---particularly TorChat \cite{wiki:torchat}, Tor Messenger \cite{wiki:tormessenger}, and Ricochet \cite{wiki:ricochet}.  In all of these services, each user is identified by a unique .onion domain.

The darkweb has a notoriously high attrition rate; its sites regularly appear and disappear. This complicates our analysis because it creates dead links, links to pages which have been removed. We do not want the dead links in our datasets so we collected responsive sites only; if we discover a link to a page on domain $v$, but domain $v$ could not be reached after $>\!10$ attempts across November 2016--February 2017, we delete node $v$ and  all edges to node $v$.  Before pruning nonresponding domains, our constructed graph had 13,117 nodes and 39,283 edges. After pruning, it has  \numnodes nodes and \numedges edges (55\% and 64\% respectively). The pruned graph is the one used in the rest of this paper, which from now on we call `` the darkweb  graph''.

We note that the darkweb as defined above is different from the network described in \cite{darknet}. There, the network represents volunteer-operated nodes that could be connected in the sense that they could appear as consecutive nodes in a path through the Tor network. This is completely different from our network.


\section{Graph-theoretic Results}
Table~\ref{fig:summary1} reports summary statistics of the darkweb and www. In what follows, we discuss these and other statistics.

\begin{table}[hbt]
\centering

\begin{tabular}{l r r} \toprule
Measure & www \cite{lehmberg2014graph} & darkweb \\
\midrule
    \textbf{Num nodes} & 43M & \numnodes \\
    \textbf{Num edges} & 623M & \numedges \\
    \textbf{Prop. connected pairs} & $\sim\!\! 42\%$ & 8.11\%  \\
    \textbf{Average SPL} & \muSPLwww & \muSPLdw \\
    \textbf{Edges per node} & 14.5 & 3.50 \\
    \textbf{Network diameter}* & 48 & 5 \\
    \textbf{Harmonic diameter} & $\sim\!\!9.55$ & 232.49 \\
\bottomrule
\end{tabular}

\caption{Summarized network level properties between the www and the darkweb.  Asterisk for the entries requiring conversion to an undirected graph.}
\label{fig:summary1}
\end{table}

\subsection{Degree distribution}
\begin{figure*}[h!bt]
	\centering
	\subfloat[darkweb in-degree]{ \includegraphics[width=0.5\textwidth]{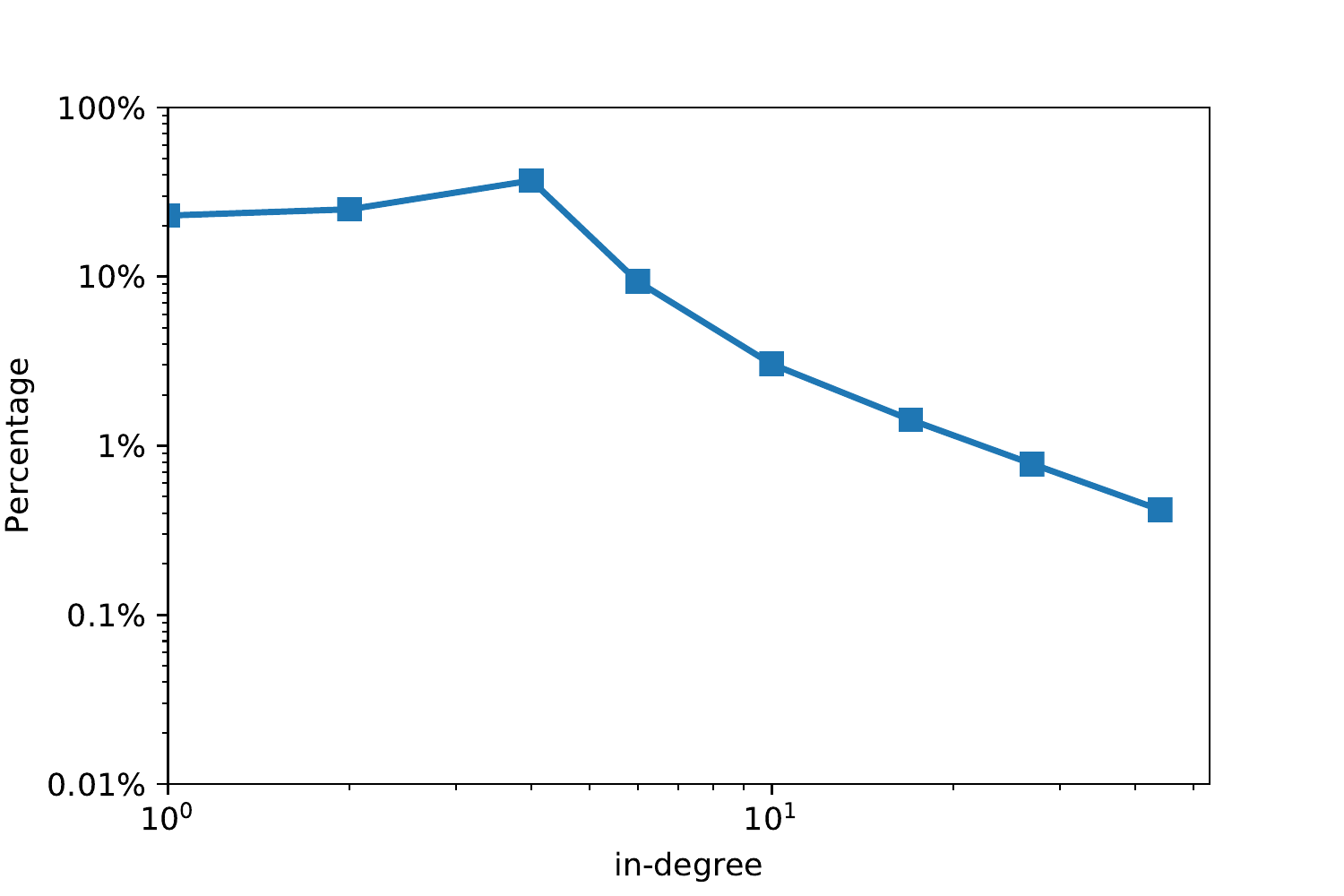} \label{fig:dw_id} }
	\subfloat[darkweb out-degree]{ \includegraphics[width=0.5\textwidth]{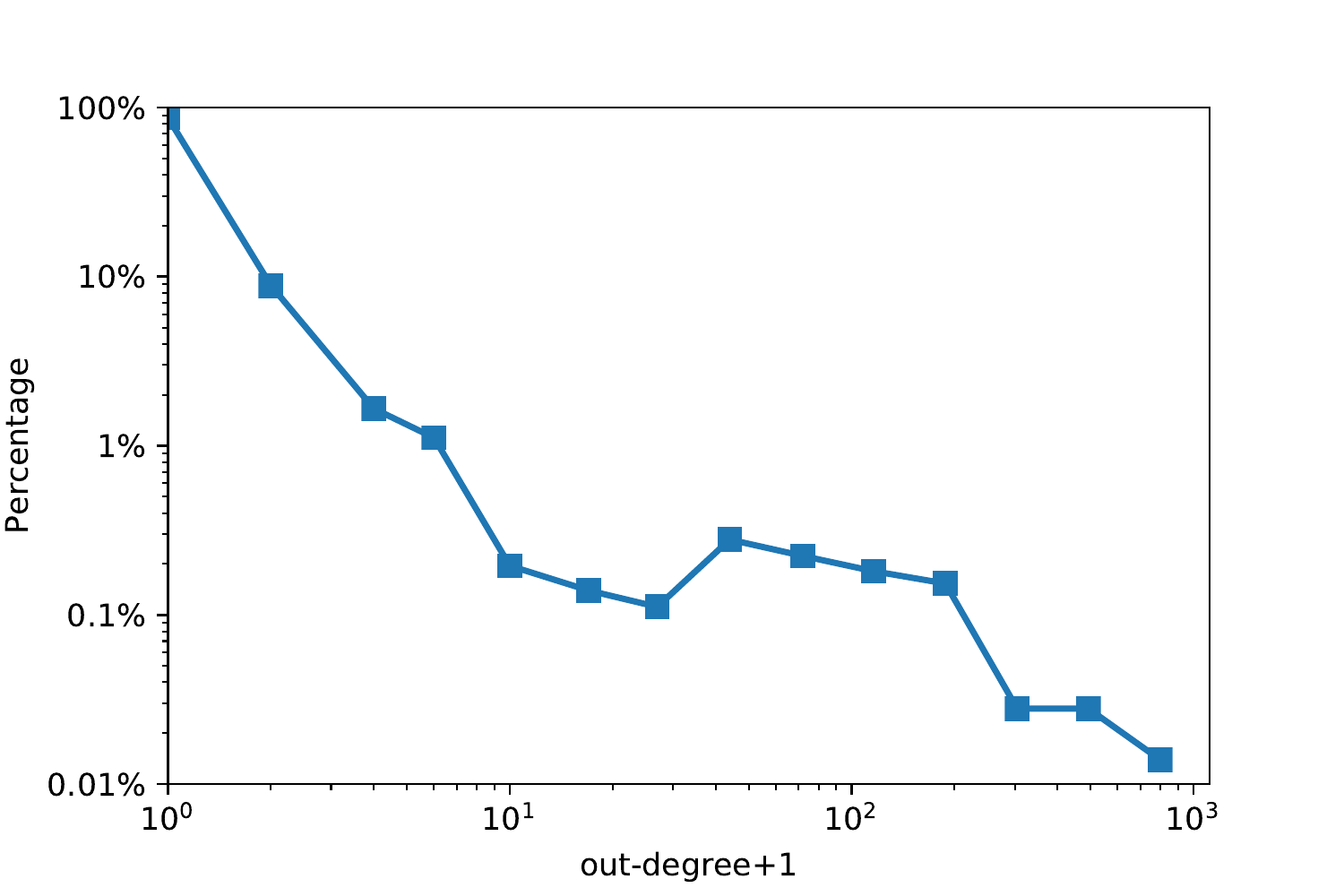} \label{fig:dw_od} }
    \\
	\subfloat[darkweb pagerank]{ \includegraphics[width=0.5\textwidth]{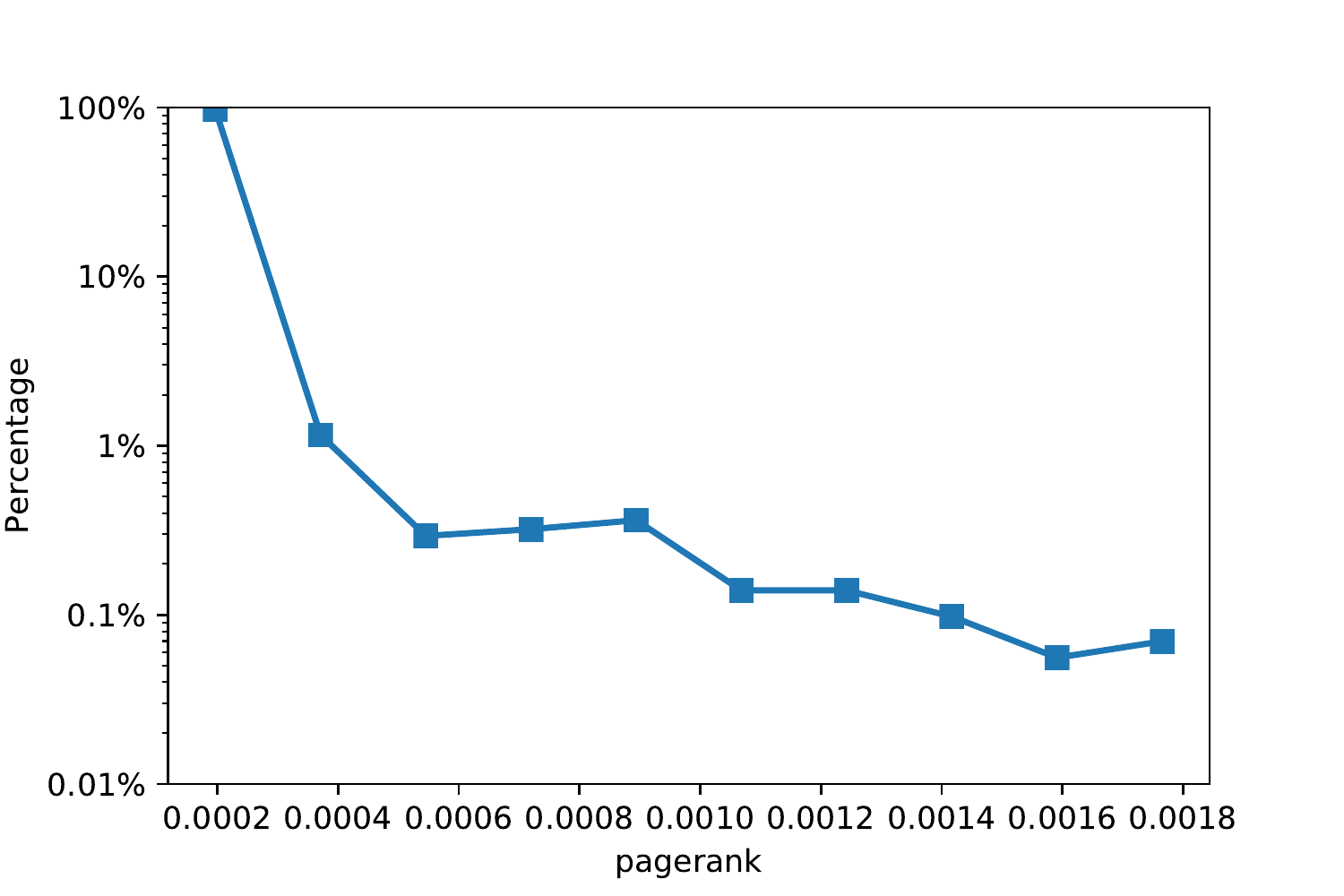} \label{fig:dw_pr} }
	\subfloat[darkweb edgeweights]{ \includegraphics[width=0.5\textwidth]{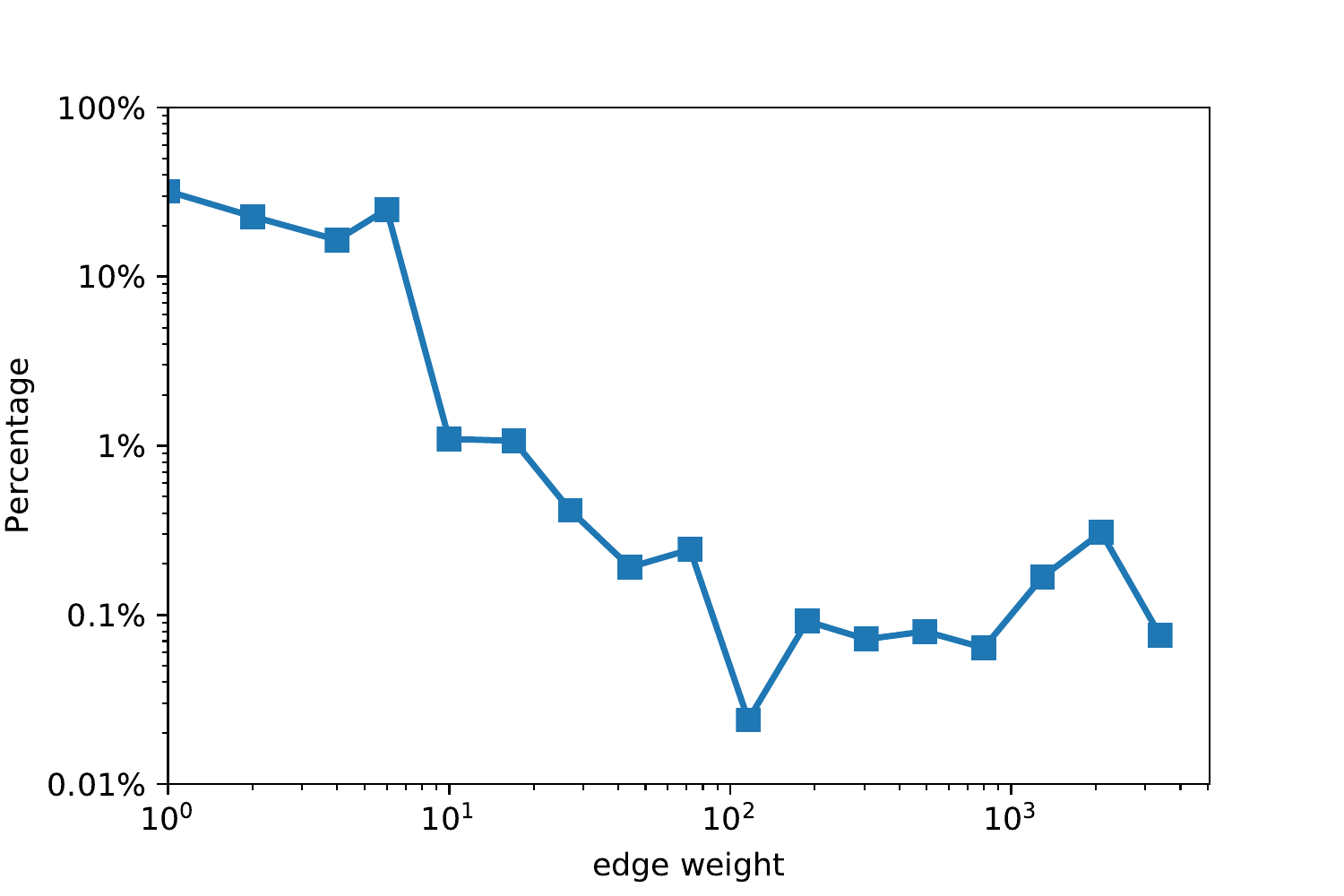} \label{fig:dw_ew} }	
	\caption{The distribution of the in-degree, out-degree, pagerank, and edgeweights. In (c) we exclude the three domains with the highest pagerank because they are such extreme outliers.  For all plots with a log-scale axis, we follow following \cite{lehmberg2014graph,meusel2015graph} to use the Fibonacci binning from \cite{vigna13fibonacci}.}
	\label{fig:degree_and_pr}
\end{figure*}

We begin with statistics on degree distributions, reported in Figure~\ref{fig:degree_and_pr}. Panels (a) and (b) show the in and out degree distributions of the darkweb resemble power laws\footnote{ Unfortunately however, we were able to confirm the power laws quantitatively. Following \cite{clauset2009power}, which describes how to fit to power laws rigorously, we tried to fit the degree distribution data using the python package \texttt{plfit}, but the fitting failed on account of there not being enough data (the range of the data didn't cover enough orders of magnitude.)}, just like the www, but with one crucial difference: the location of the $y$-intercept. In (a) we see $\sim\! 30\%$ of domains have \emph{exactly one} incoming link ($k_{in} = 1$), with $62\%$ come from one of the five largest out-degree hubs. In (b), we see a whopping \percentzeroOD of sites \emph{do not link to any other site} ($k_{out} = 0$)! -- these are the dark silos that we mentioned in the abstract. 

These findings tell us the darkweb is a sparse hub-and-spoke place. The bulk of its sites live in seclusion, not wanting to connect with the outside world. Panel (c) confirms this picture by showing the vast majority of pages have low pagerank (and so are isolated). Panel (d) shows that when a site does link to another, $32\%$ of the time it's only a single page linking out. 

As we show in the next section, this siloed structure is not shared by many social networks.

\subsection{Bow-tie decomposition}
A useful way to describe directed graphs is via \emph{bow-tie decomposition}, where the nodes are divided into six disjoint parts \cite{broder2000graph}:
\begin{enumerate}
    \item \textsc{CORE} --- Also called the ``Largest Strongly Connected Component''.  It is defined as the largest subset of nodes such that there exists a directed path (in both directions, from $u \rightarrow \cdots \rightarrow v$ as well as $v \rightarrow \cdots \rightarrow u$) between every pair of nodes in the set.
    \item \textsc{IN} --- The set of nodes, excluding those in the \textsc{CORE}, that are ancestors of a \textsc{CORE} node.
    \item \textsc{OUT} --- The set of nodes, excluding those in the CORE, that are descendants of a \textsc{CORE} node.
    \item \textsc{TUBES} --- The set of nodes, excluding those in the \textsc{CORE}, \textsc{IN}, and \textsc{OUT}, who have an ancestor in \textsc{IN} as well as a descendant in \textsc{OUT}.
    \item \textsc{TENDRILS} --- Nodes that have an ancestor in \textsc{IN} but do not have a descendant in \textsc{OUT}.  Also, nodes that have a descendant in OUT but do not have an ancestor in \textsc{IN}.
    \item \textsc{DISCONNECTED} --- Everything else.
\end{enumerate}

\begin{figure*}[hbt]
\centering

\hbox{\hspace{5em} \includegraphics[width=0.9\textwidth]{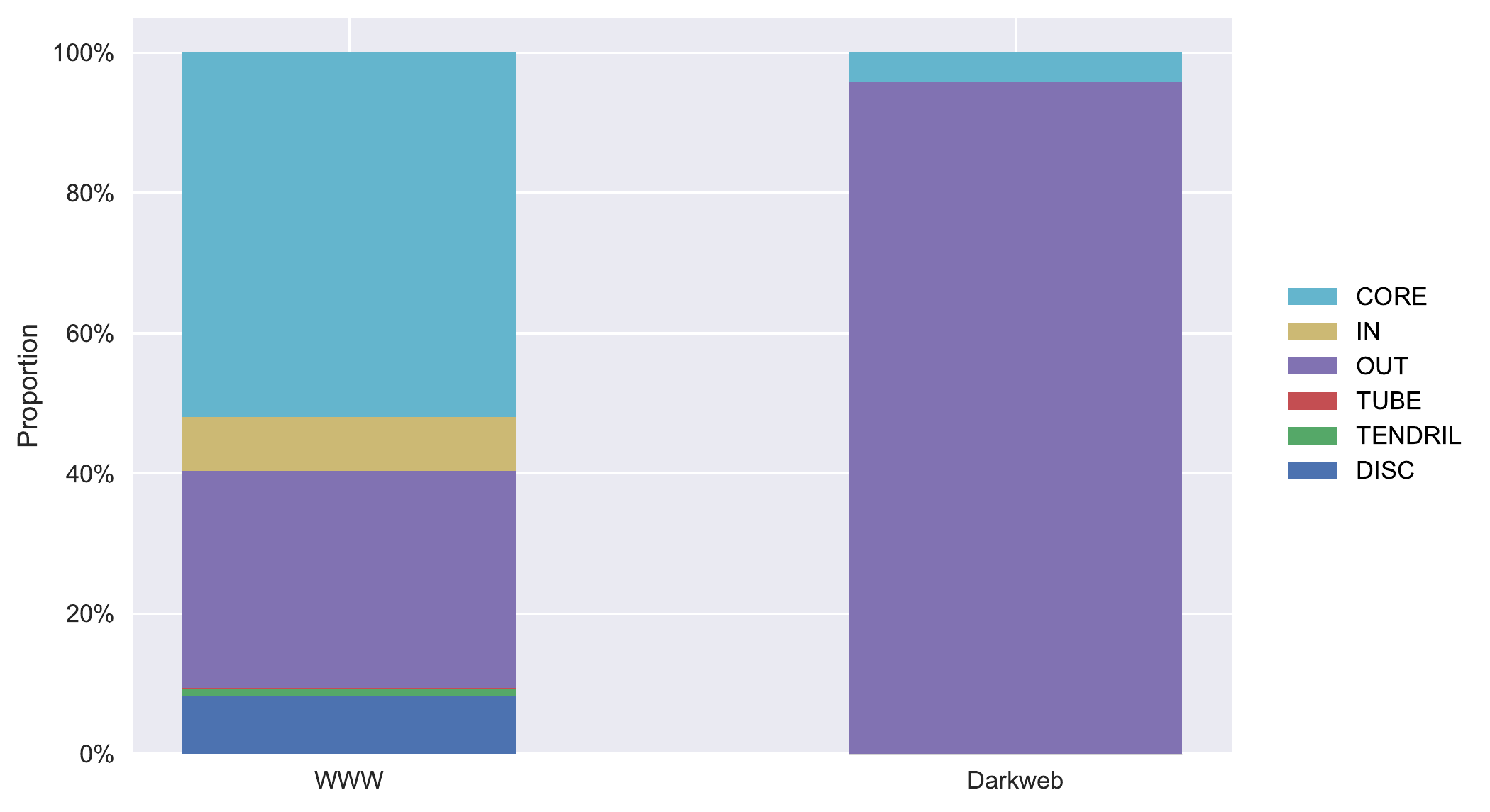} \label{fig:bowtie_stacked} }

\vspace{1.5em}

\begin{tabular}{l | r r r r} \toprule
& \multicolumn{2}{c}{World-Wide-Web} & \multicolumn{2}{c}{Darkweb} \\
\midrule
	CORE         & 22.3M & 51.94\% & 297 & 4.14\% \\
	IN           & 3.3M  &  7.65\% & 0 & 0.0\% \\
	OUT          & 13.3M & 30.98\% & 6881 & 95.86\% \\
	TUBES        & 17k   &   .04\% & 0 & 0.0\% \\
	TENDRILS     & 514k  &   1.2\% & 0 & 0.0\% \\
	DISCONNECTED & 3.5M  &   8.2\% & 0 & 0.0\% \\
\bottomrule
\end{tabular}

\caption{Bow-tie decompositions of the www and darkweb. The figures for the www were taken from \cite{lehmberg2014graph}. }
\label{fig:bowtie}
\end{figure*}

Figure~\ref{fig:bowtie} compares the bowtie decompositions of the darkweb and www. The www numbers are taken from \cite{Serrano2007,lehmberg2014graph,meusel2015graph}. We chose these works because of the size of their crawls and the rigor of their analyses. Notice the www has each of one the 6 components of the bow-tie decomposition, whereas the darkweb only has a \textsc{CORE} and an \textsc{OUT} component. Moreover, the OUT component of the darkweb contains $\sim 96\%$ of the mass (these are the dark silos), leaving the CORE with just $\sim 4 \%$. This is unusual for a social network; most have large COREs. The www's CORE has $> 50 \%$ of the mass, while the cores of Orkut, YouTube, Flickr \cite{mislove2007measurement} and Twitter \cite{garcia2017understanding} are even larger. In this sense, the darkweb is a  social network anomaly.

\subsection{Diameter analysis}
\begin{figure*}[h!bt]
	\centering
	\subfloat[\hspace{1em}word wide web\newline $\mu=\muSPLwww$]{ \includegraphics[width=0.33\textwidth]{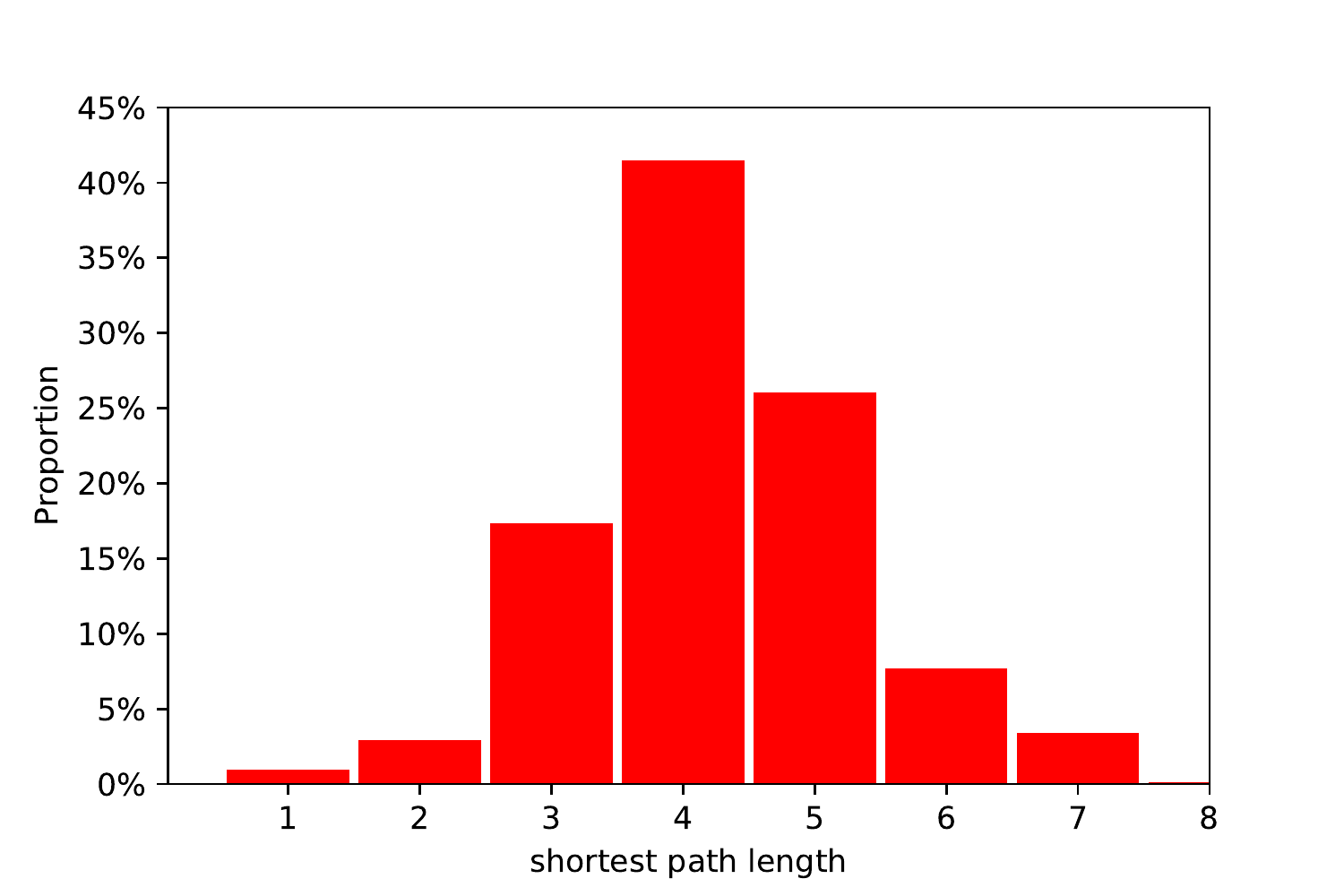} \label{fig:www_spl} }
	\subfloat[\hspace{2em}darkweb\newline $\mu=\muSPLdw$]{ \includegraphics[width=0.33\textwidth]{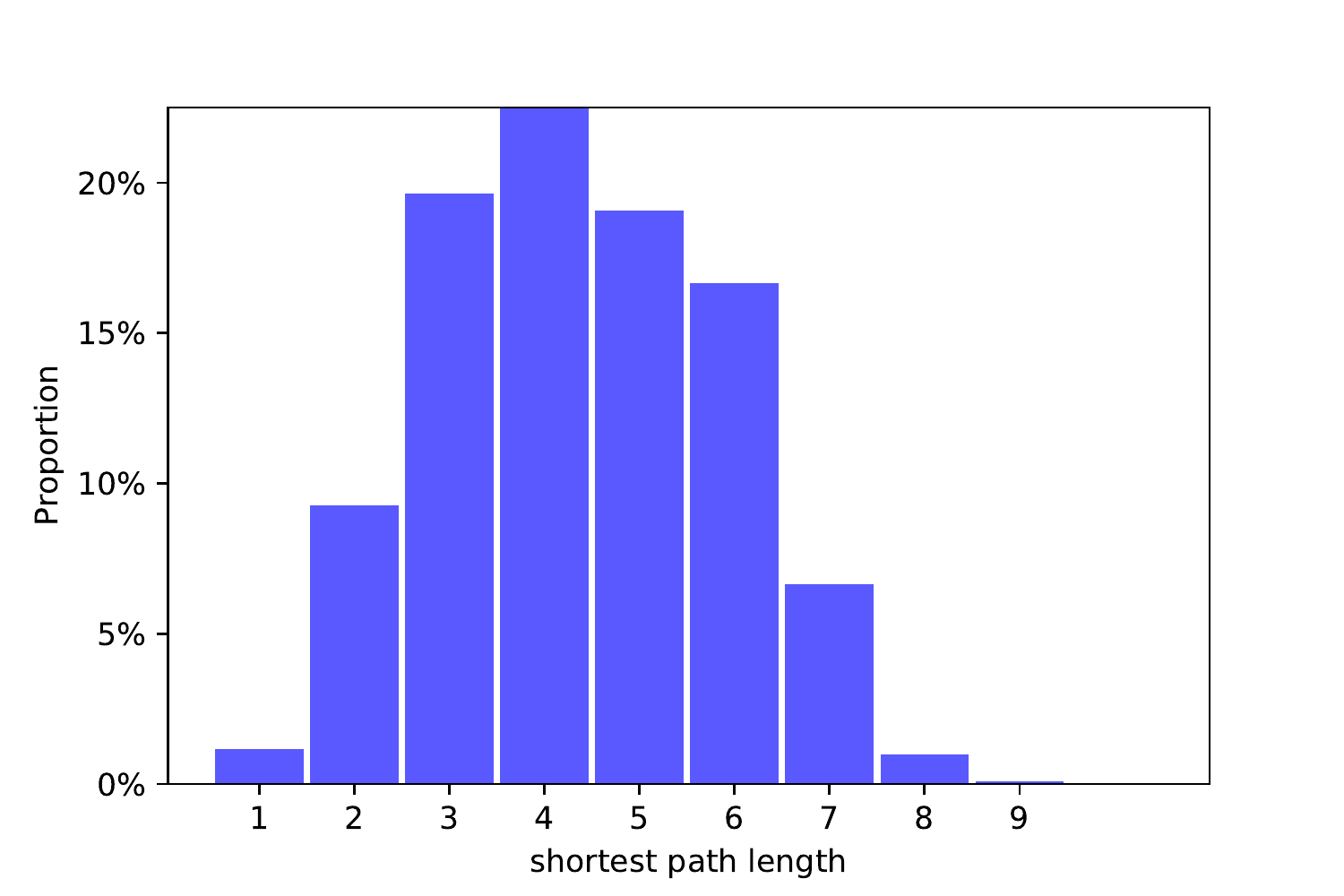} \label{fig:dw_spl} }
	\subfloat[\hspace{1em}darkweb \textsc{CORE}\newline $\mu=\muSPLcore$]{ \includegraphics[width=0.33\textwidth]{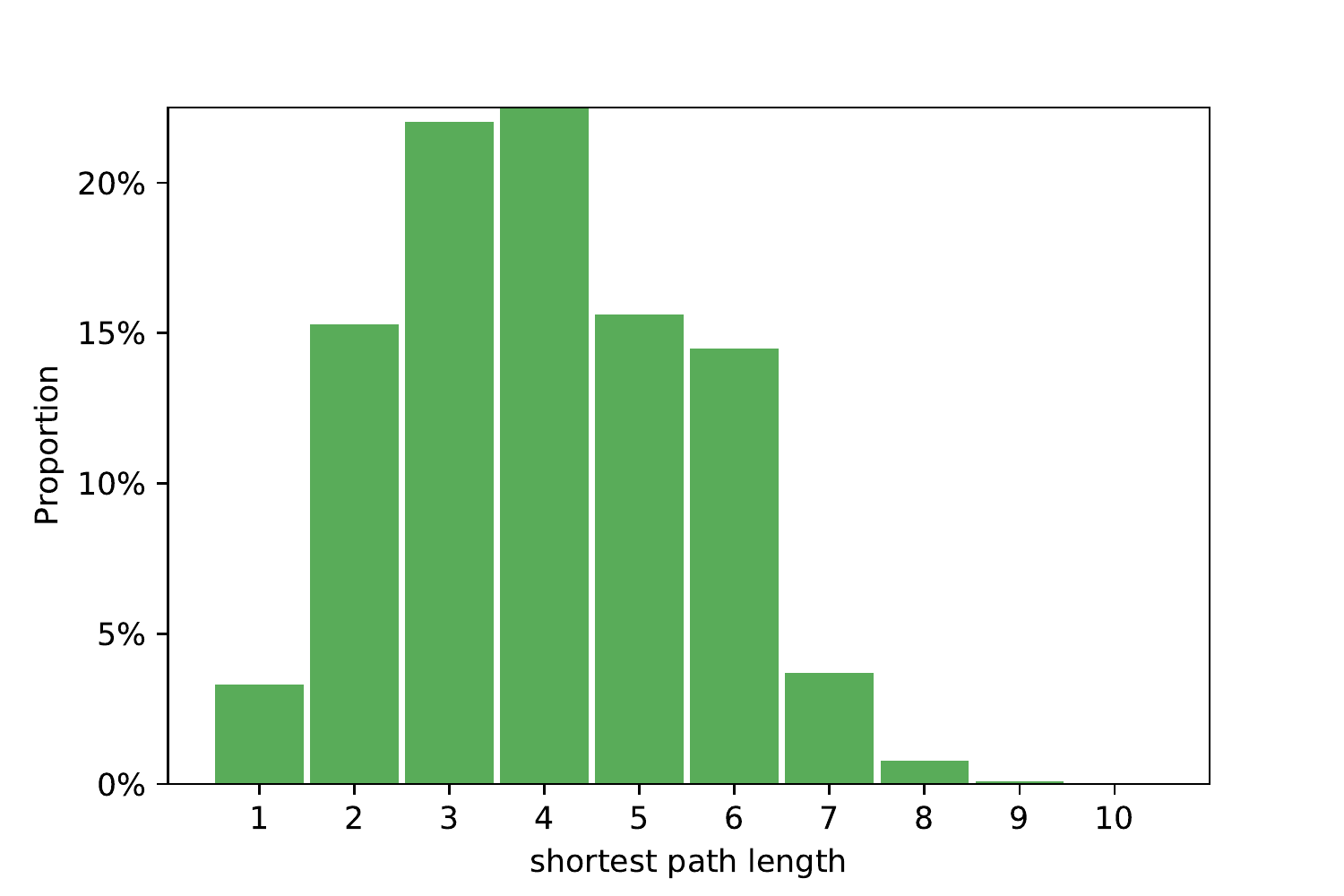} \label{fig:dw_core_spl} }	
	\caption{Comparing shortest path lengths between the world-wide-web and the darkweb considering directed edges.  Whereas in the www $56.65\%$ of node pairs have have $\infty$ path-length (no path connecting them), in the darkweb $91.89\%$ of node-pairs have no path connecting them.  Moreover, even within that $8.11\%$ of pairs with a directed path between them, the darkweb's average SPL ($\mu = \muSPLdw$) is \emph{higher} than that of the www ($\mu = \muSPLwww$).}
	\label{fig:spl}
\end{figure*}

Next we examine the darkweb's internal connectivity by computing the shortest-path-length (SPL) between nodes. Figure~\ref{fig:spl}(a) and (b) shows the SPL's for the www and darkweb. For all pairs of nodes $\{u,v\}$ in the darkweb, only 8.11\% are connected by a directed path from $u \rightarrow \cdots \rightarrow v$ or $v \rightarrow \cdots \rightarrow u$.  This is drastically lower than the $\sim\!\!43.42\%$ found in the www \cite{lehmberg2014graph}.  We again attribute this to the low out-degree per \ref{fig:degree_and_pr}.  Of the connected pairs, the darkweb's average shortest path length is \muSPLdw compared to the \muSPLwww in the world-wide-web \cite{lehmberg2014graph}.  It's surprising to see a graph as small as the darkweb have a higher mean SPL than the entire world-wide-web, and is a testament to how sparse the darkweb graph really is. Figure~\ref{fig:spl}(c) plots the distribution of SPLs for the \COREsize nodes of the \textsc{CORE}. To our surprise, the mean SPL within the \textsc{CORE} is \muSPLcore, only $9\%$ less than the entire darkweb.  From this we conclude the \textsc{CORE} is not densely interconnected.

\subsection{Robustness and Fragility}
\begin{figure*}[h!bt]
	\centering

	\subfloat[removing nodes by lowest in-degree]{ \includegraphics[width=0.45\textwidth]{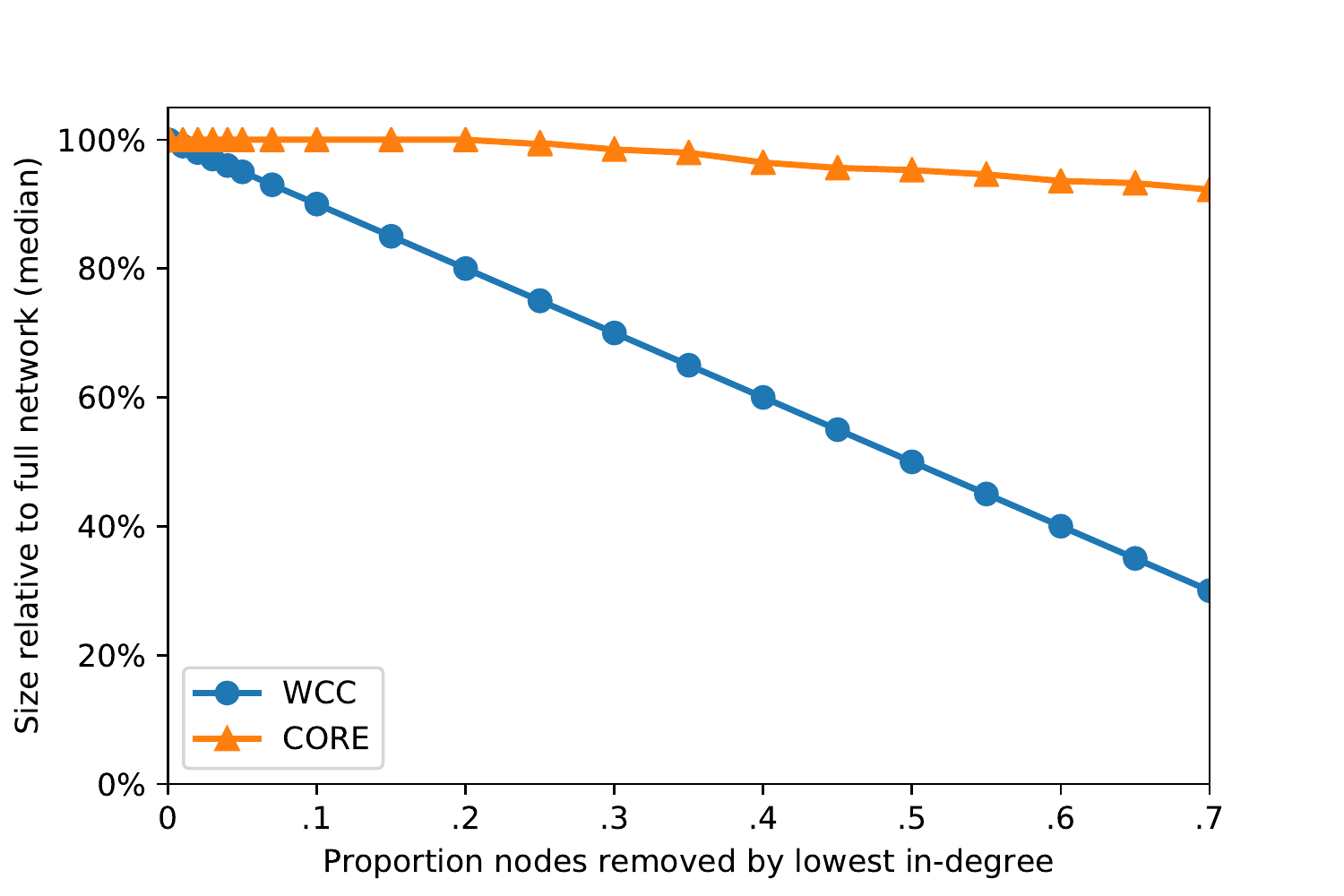} \label{fig:dw_removing_indgree_lowest} }
	\subfloat[removing nodes by highest in-degree]{ \includegraphics[width=0.45\textwidth]{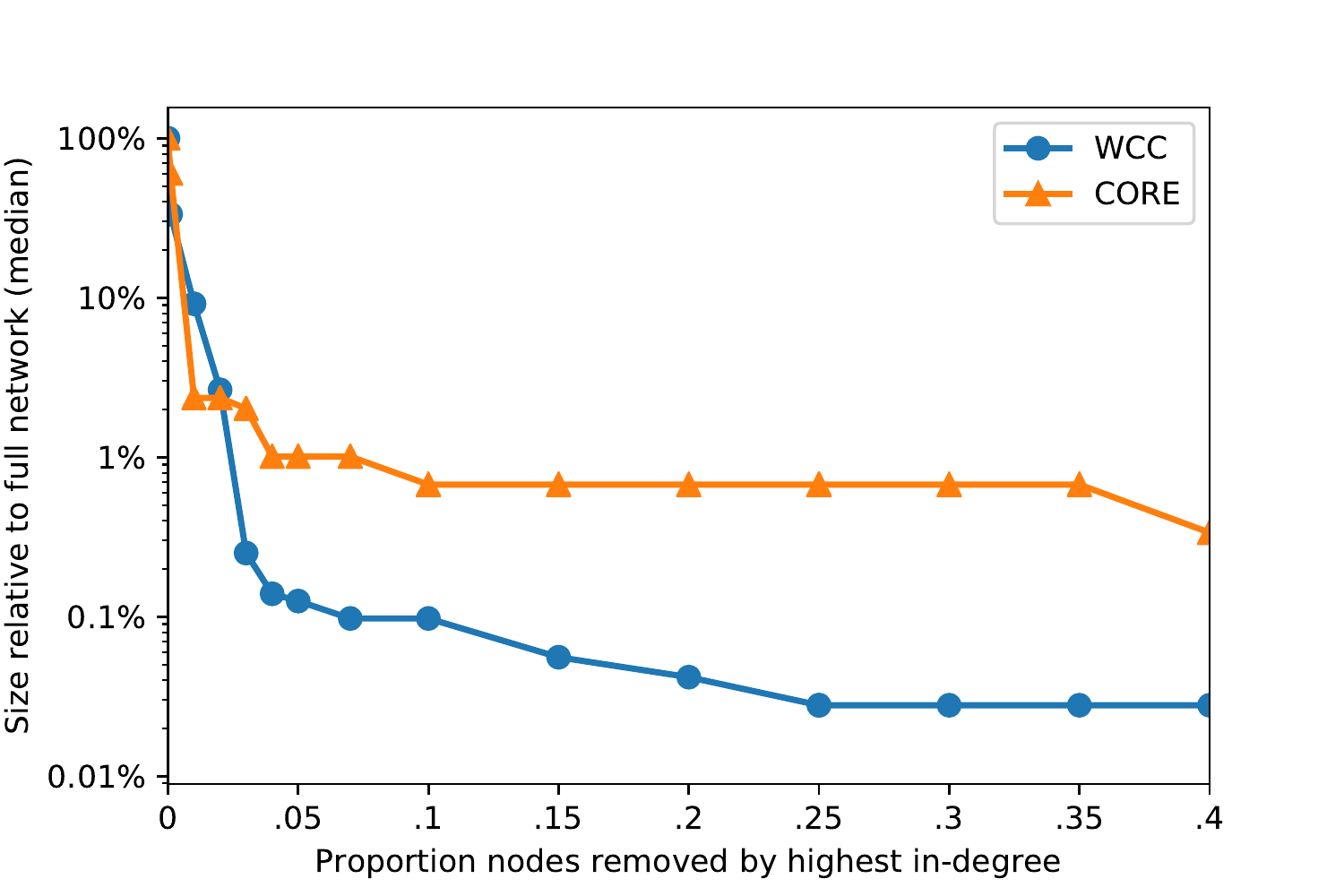} \label{fig:dw_removing_indgree_highest} }

	\subfloat[removing nodes by lowest harmonic centrality]{ \includegraphics[width=0.45\textwidth]{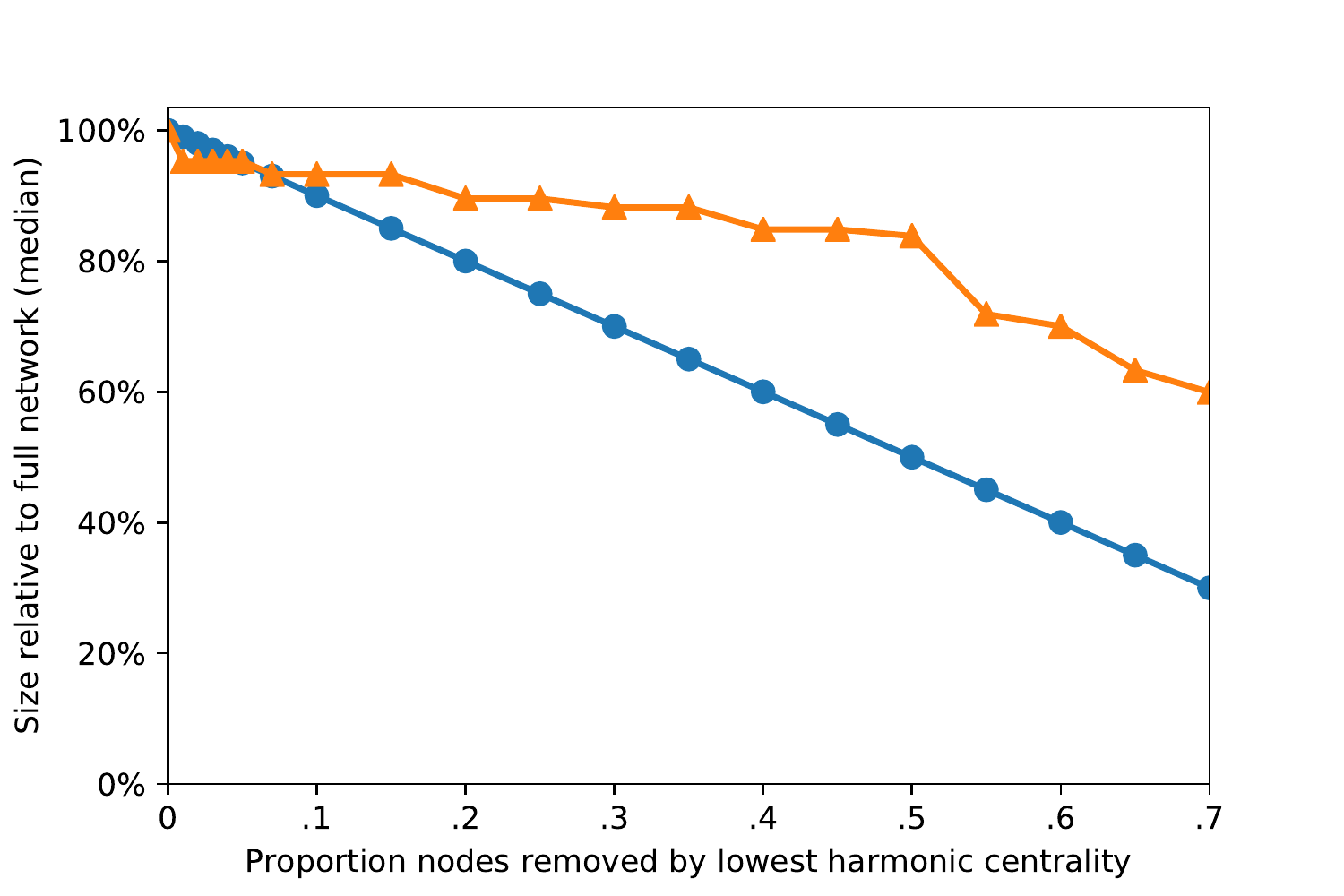} \label{fig:dw_removing_harmonic_lowest} }
	\subfloat[removing nodes by highest harmonic centrality]{ \includegraphics[width=0.45\textwidth]{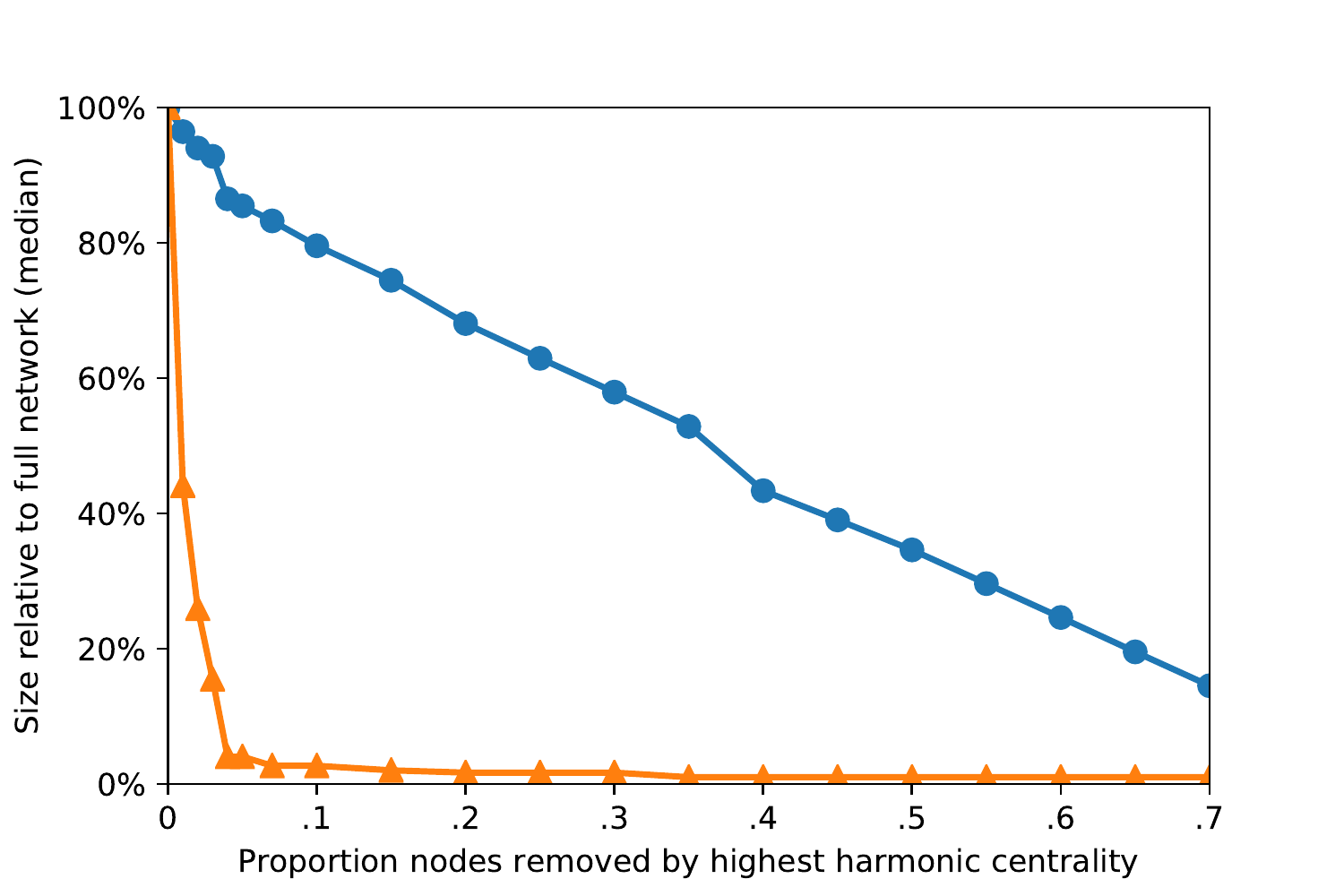} \label{fig:dw_removing_harmonic_highest} } \\

	\subfloat[removing nodes by lowest pagerank]{ \includegraphics[width=0.45\textwidth]{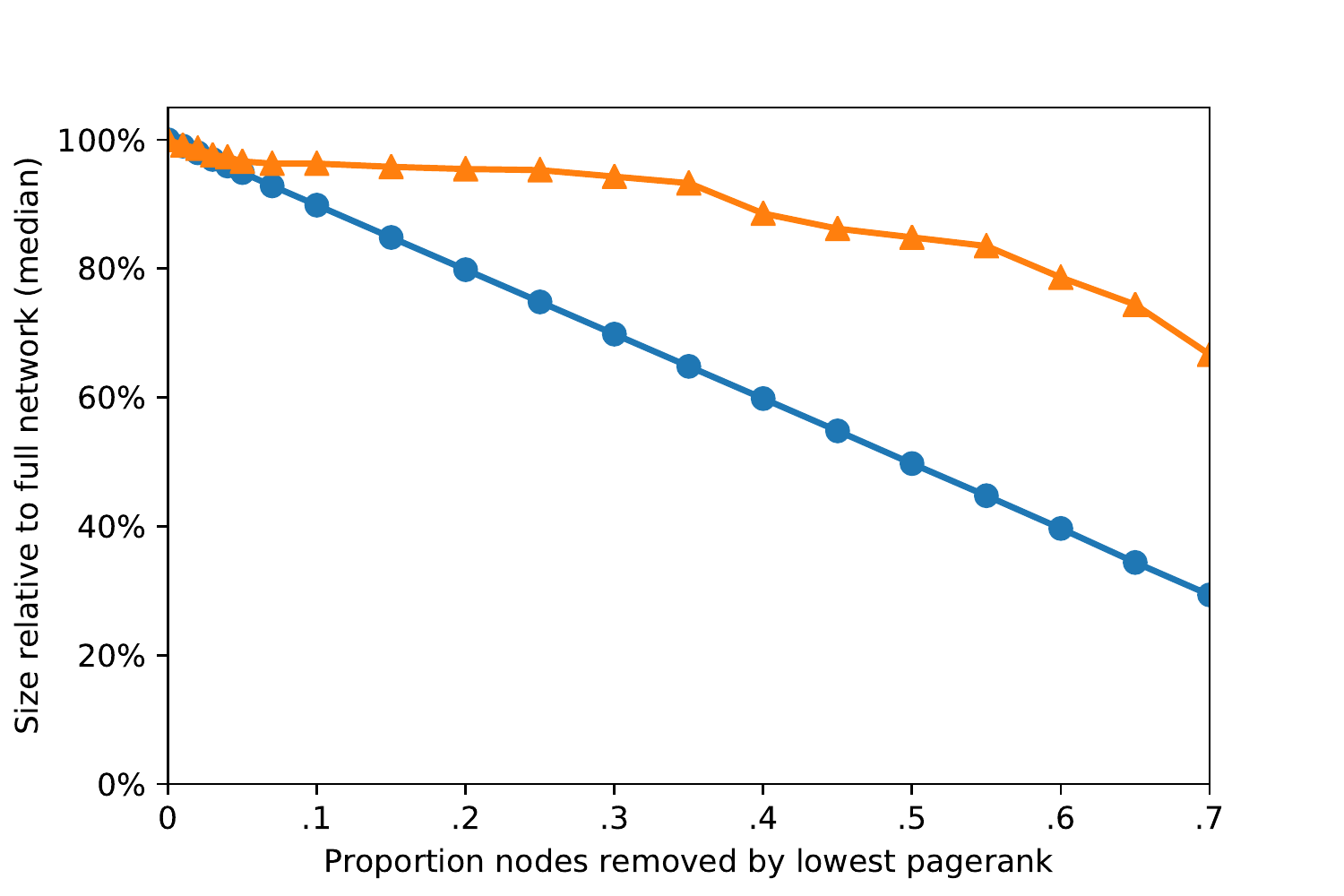} \label{fig:dw_removing_pagerank_lowest} }
	\subfloat[removing nodes by highest pagerank]{ \includegraphics[width=0.45\textwidth]{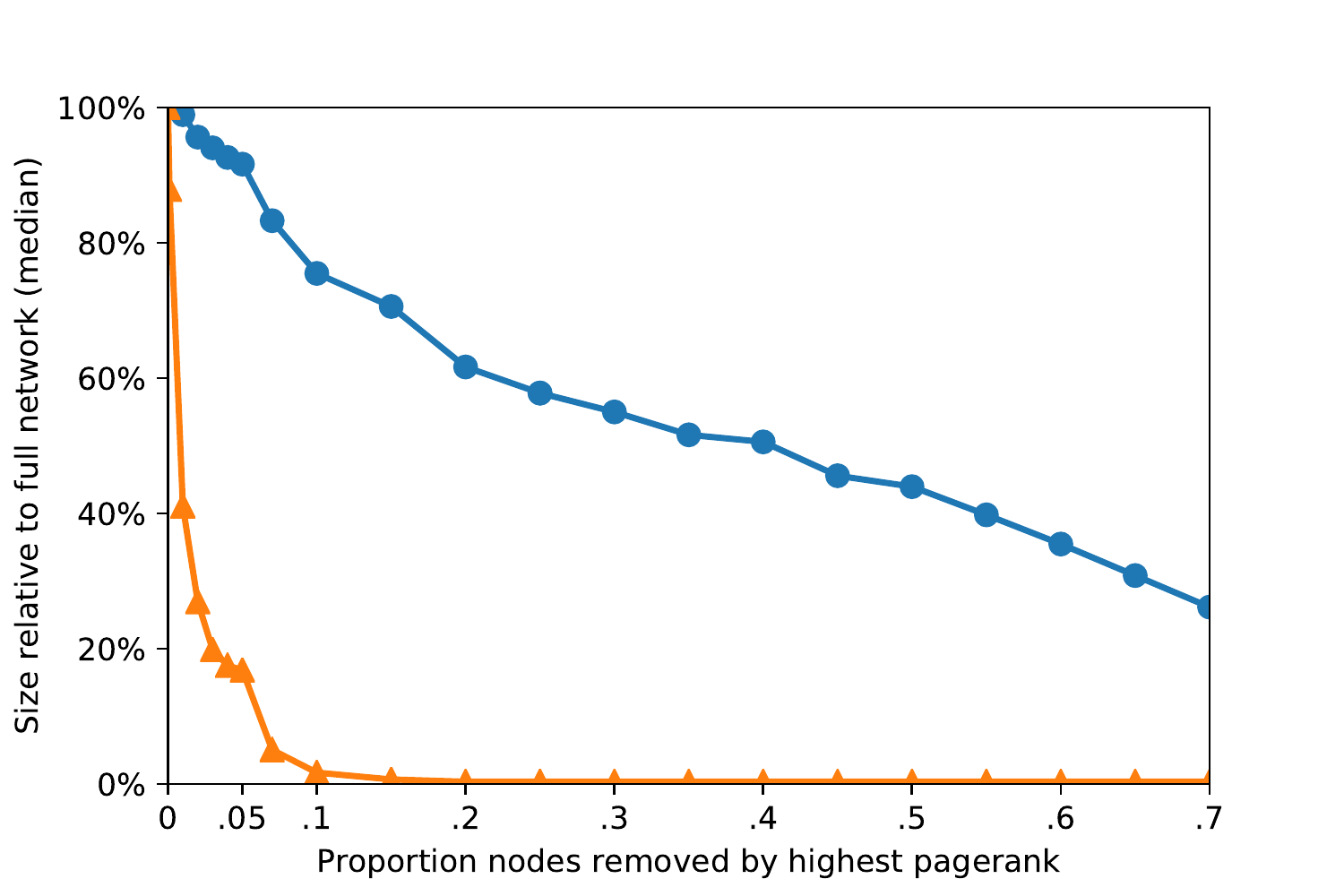} \label{fig:dw_removing_pagerank_highest} }

	\caption{Deleting nodes from the darkweb graph and seeing how quickly the WCC and \textsc{CORE} disintegrate.  In all plots, we shuffled the order of nodes with the same value until reaching stable statistics, e.g., in \ref{fig:dw_pr}, $98\%$ of nodes are tied for the lowest pagerank; so when removing only $10\%$ of the nodes (e.g., \ref{fig:dw_removing_pagerank_lowest}), it's ambiguous which nodes should be deleted first.  So in our analysis we shuffled the ordering of the nodes with the same value and recomputed sizes of the WCC/\textsc{CORE} until the median was stable.}
	\label{fig:knockouts}
\end{figure*}

Does the peculiar structure of the darkweb make it resilient to attack? Figure~\ref{fig:knockouts} shows it does not. As seen, the entire network (WCC) as well as the \textsc{CORE} quickly disintegrates under node removal.  In Figures~\ref{fig:knockouts}(a) and (b) we see the familiar resistance to random failure yoked with fragility to targeted attacks, in keeping with the findings of \cite{bollobas2004robustness}.  Figures~\ref{fig:knockouts}(b) shows that, unlike the www \cite{lehmberg2014graph}, the WCC is \emph{more susceptible to high in-degree deletions than the \textsc{CORE}}.  This elaborates the view from Figures~\ref{fig:knockouts}(c) that the \textsc{CORE} is -- in addition to not being strongly interconnected -- also not any kind of high in-degree nexus.

Figures~\ref{fig:knockouts}(c) and (d) show the breakdown when removing central nodes.  In (c), the CORE is largely unaffected by low centrality deletions. In Figure~\ref{fig:knockouts}(c) we see that although the \textsc{CORE} isn't disproportionately held together by high in-degree nodes, it is dominated by very central nodes.

Comparing Figures~\ref{fig:knockouts}(b) and (f) we see the \textsc{CORE} relative to the entire network consists of more high-pagerank nodes than high in-degree nodes.  This implies \textsc{CORE} nodes are not created by their high-indegree (\ref{fig:dw_removing_indgree_highest}), but by their high centrality, amply corroborated by Figures~\ref{fig:knockouts}(c) and (d).  Likewise, Figures~\ref{fig:knockouts}(e) recapitulates \ref{fig:dw_removing_indgree_lowest}, that nodes with especially low in-degree or centrality are, unsurprisingly, not in the \textsc{CORE}.

\subsection{Reciprocal Connections}
\begin{figure*}
\centering 
\includegraphics[height=3cm]{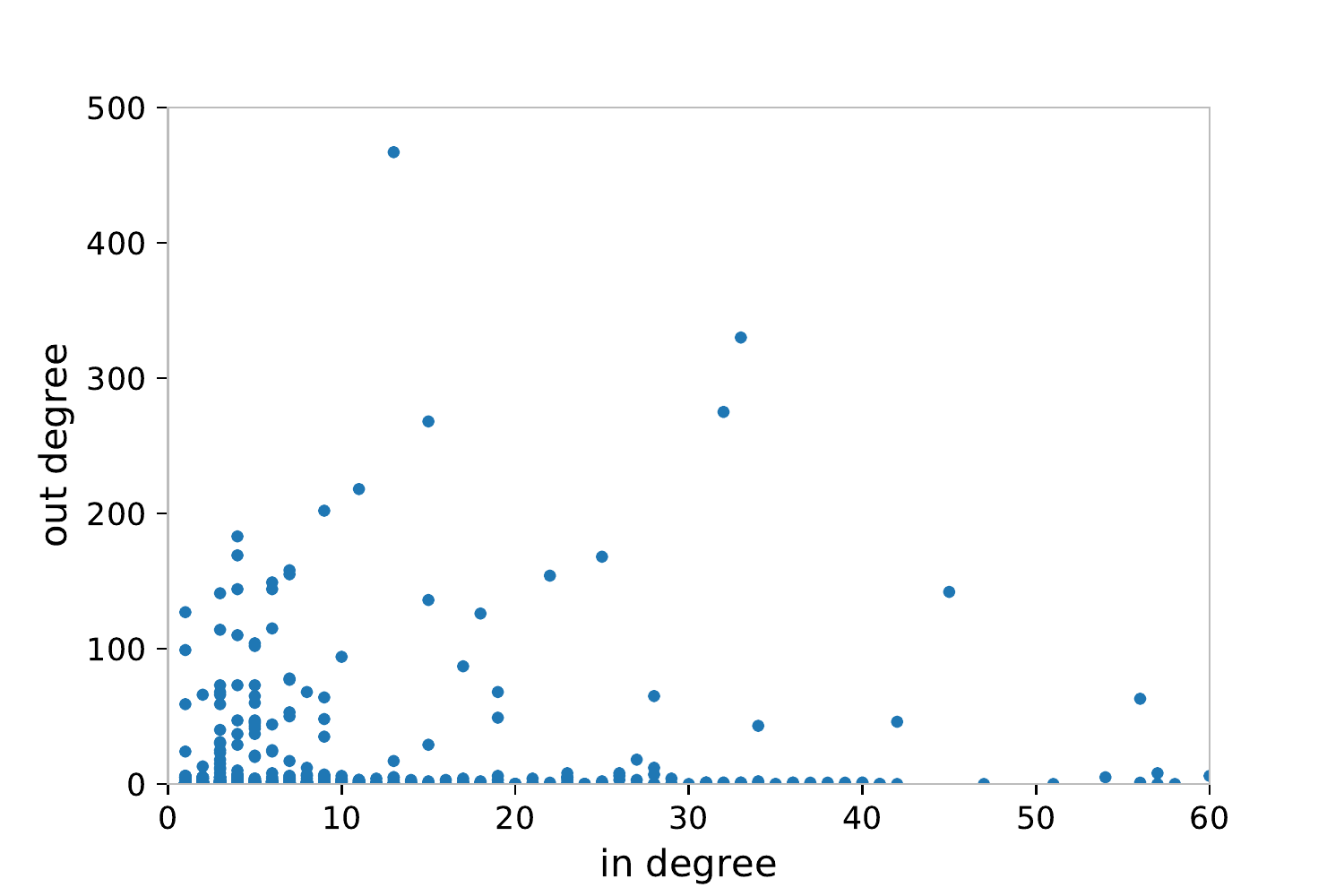}\quad
\includegraphics[height=3cm]{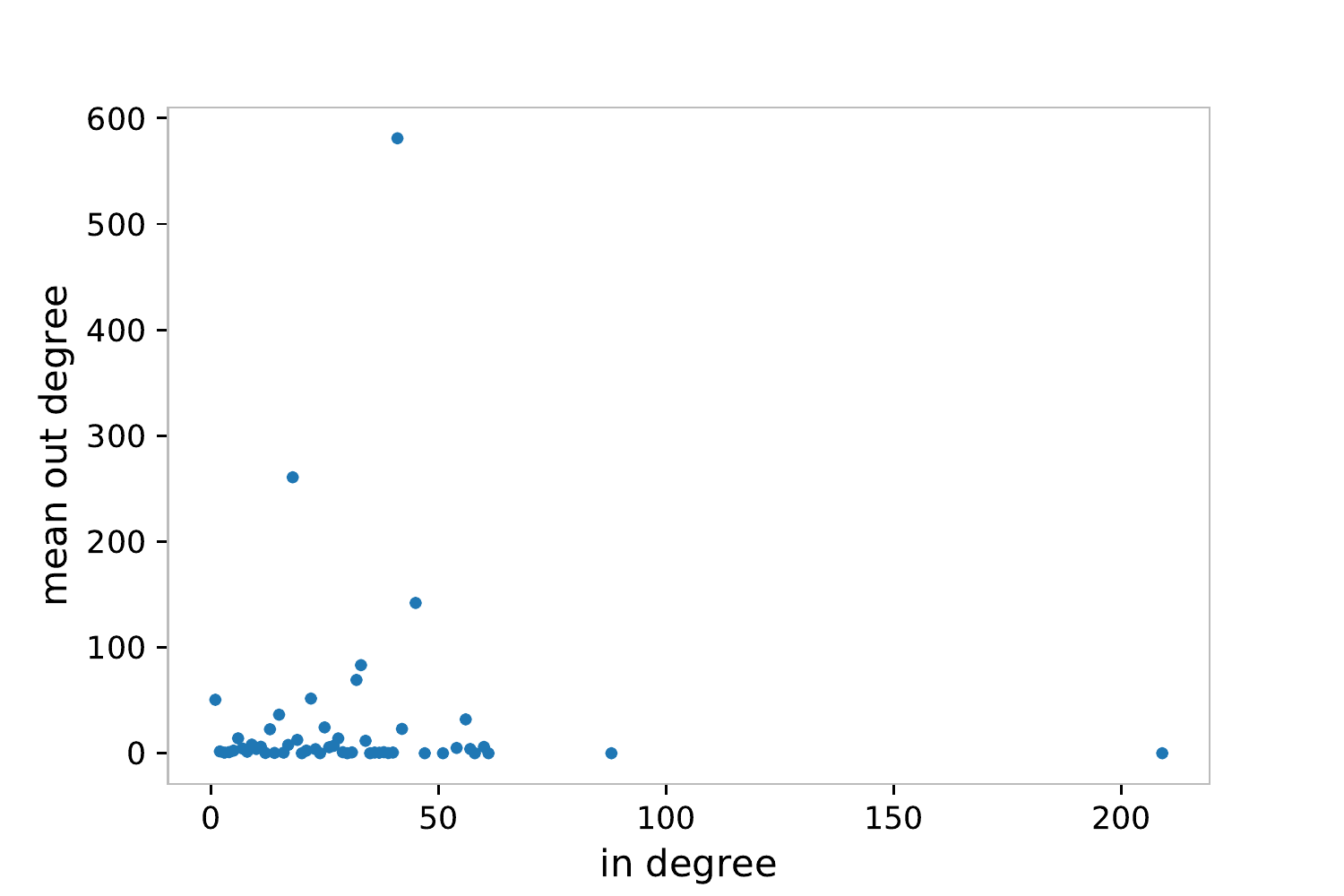}\par\medskip
\includegraphics[height=3cm]{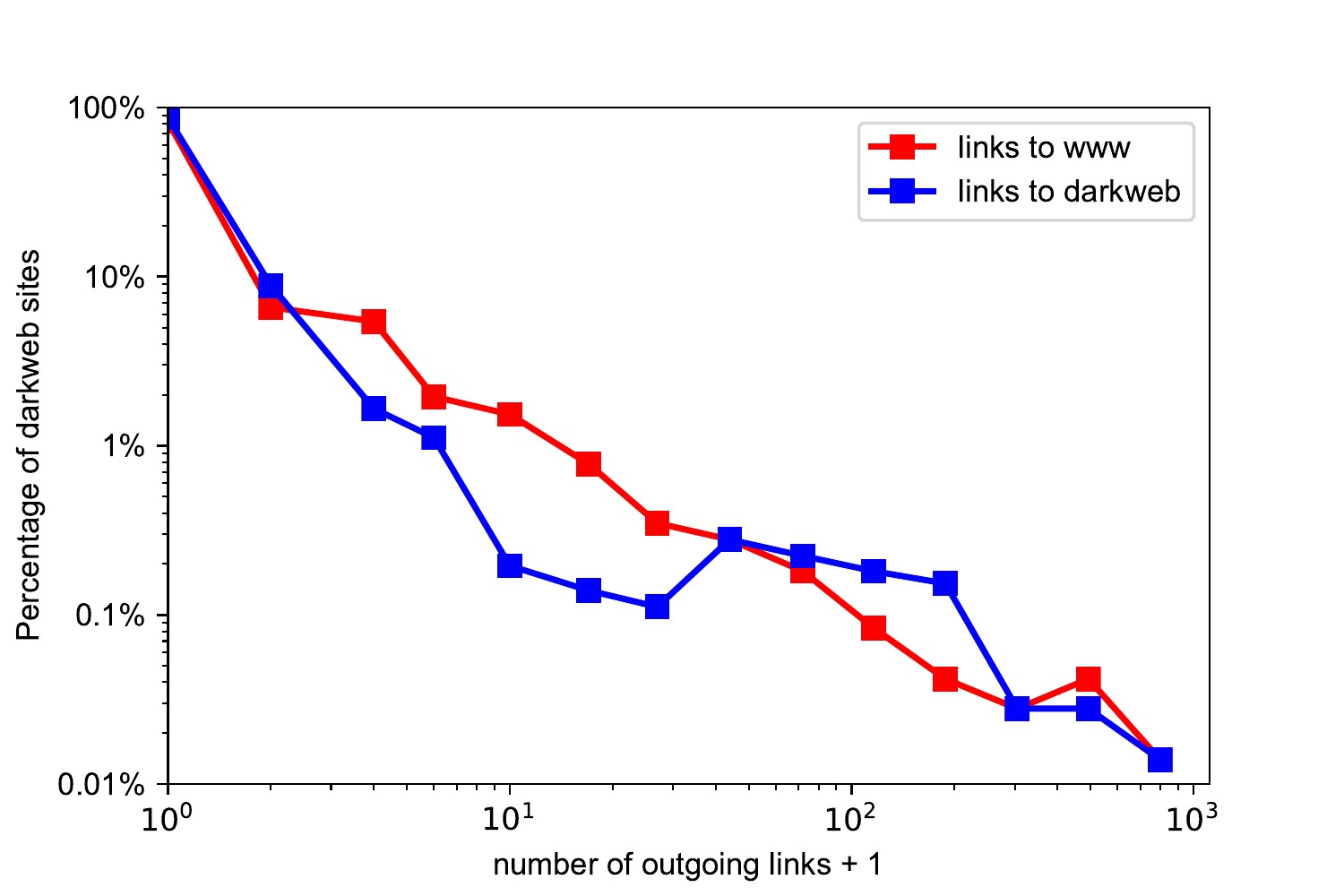}
\caption{(a): Scatter plot of $k_{in}$ and $k_{out}$. (b) Plot of averaged $k_{out}$ versus $k_{in}$. (c) Comparing the rates of the darkweb sites linking to the www versus linking to other darkweb sites. They are essentially the same.}
\label{fig:outdeg_as_func_of_indeg}
\end{figure*}

The authors of \cite{Serrano2007} stress the importance of reciprocal connections in maintaining the www's graph properties.  We compute two of their measures.  First, we compute \cite{Serrano2007}'s measure $\frac{\langle k_{in}k_{out} \rangle }{ \langle k_{in} \rangle \langle k_{out} \rangle } = \frac{ \mathbb{E}[ k_{in} k_{out} ] }{ \mathbb{E}[ k_{in} ] \mathbb{E}[ k_{out} ] }$, to quantify in-degree and out-degree's deviation from independence.  For the darkweb, we arrive at $\frac{\langle k_{in}k_{out} \rangle}{\langle k_{in} \rangle \langle k_{out} \rangle}=3.70$.  This is in the middle of the road of prior estimates of the www, and means that the out-degree and in-degree are positively correlated.  For greater clarity, we also plot the average out-degree as a function of the in-degree, given as,
\begin{equation}
\langle k_{out}(k_{in}) \rangle = \frac{1}{N_{k_{in}}} \sum_{i \in \Upsilon(k_{in})} k_{out,i} \; ,
\label{eq:serrano}
\end{equation}
which is simply ``For all nodes of a given in-degree, what is the mean out-degree?''.  The results are depicted in Figure~\ref{fig:outdeg_as_func_of_indeg}. In short, in the darkweb there's no obvious pattern to the relationship between in-degree and out-degree.

\subsection{Network growth model}
Are standard network growth models able to capture the topology of the darkweb? Here we show a generalized preferential attachment model approximately can. In regular preferential attachment, nodes of only one type are added sequentially to a network. Here, we generalize this procedure to two types of nodes: \textit{pages} and \textit{portals}. Page nodes model the nodes in the darkweb which do not link to anyone (and so have $k_{in} = 0$). Portals model the rest of the nodes, which act as `portals' to the page nodes. The dynamics of the `page-portal' model are defined as follows.

 At $t = 0$ $N_0$ portals are arranged in a given network. At each time step after $t>0$, a new page is added to the network and $m$ different portals are chosen to link to $m$ end nodes (these end nodes can be either pages or portals). The portal which does the linking is chosen according to preferential attachment, that is, chosen with probability $\propto (1 + k_{out}^{\beta}$), where $k_{out}$ is the out degree and $\beta$ is a free parameter. The end node chosen with probability  $\propto (1 + k_{in}^{\beta})$. Notice this scheme means that only portals can link, which means page nodes forever remain with $k_{in} = 0$, as desired. The model has three free parameters $m,  \beta , N_0$. We chose $N_0 = 397$ so as to match the number of portals in our datasets (defined as all nodes with $k_{in} > 0$) and ran the model for $6242$ timesteps i.e added $6242$ page nodes) so that there were $7178$ nodes at the end of the process which again matched the dataset.
 
\begin{figure*}[!htbp]
    \centering
    \includegraphics[width=.75\linewidth]{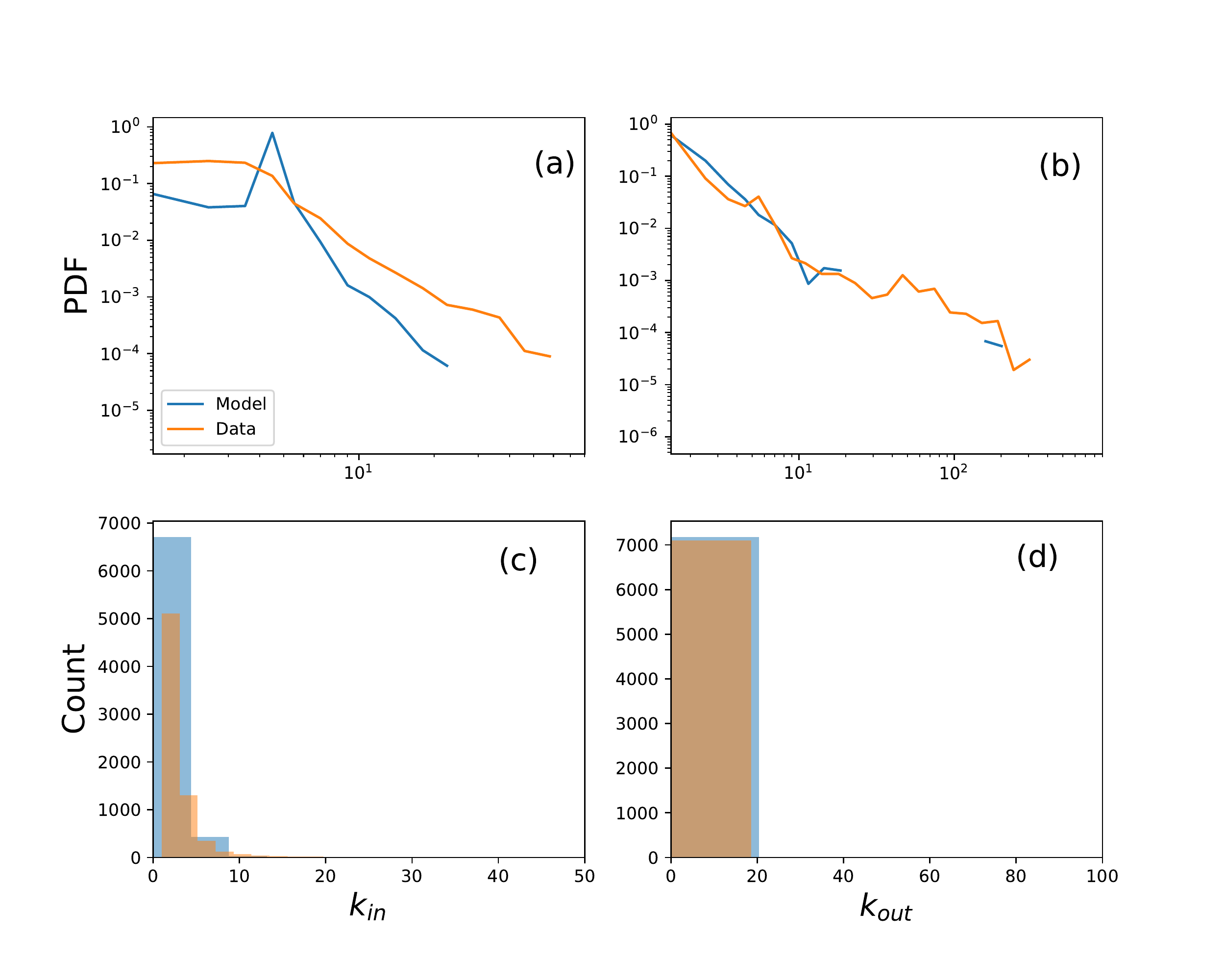}
    \caption{Comparison of in ($k_{in}$) and out ($k_{out}$) degree statistics of darkweb and page-portal model.  (a) Probability density functions of $k_{in}$ and $k_{out}$ for page and portal model. (c)-(d) Histograms of above quantities.}
    \label{alphas}
\end{figure*}

Figure~\ref{alphas} shows the page-portal model with parameters $(m,\beta)  = (4, 2)$ approximates the darkweb, the in and out degree distributions of the model approximately mimicking the data.  We report exponents of best fit, found using the powerlaw package python. But keep in mind that as mentioned earlier, there are not enough orders of magnitude in the data for the estimates of the exponents to be reliable; thus the page and portal model is intended as a first step in modeling the darkweb. The values were $(\alpha_{in}, \alpha_{out})_{data} = (3.09, 2.10)$ and $(\alpha_{in}, \alpha_{out})_{model} = (4.3 \pm 1.3 , 2.4 \pm 0.2 )$, where the model figures report means of 10 realizations and the error is given as half the range. The fitting was performed using the python powerlaw package. $\alpha_{in}$ was much more difficult to measure than $\alpha_{out}$. As much as half of realizations led to estimates $ \alpha_{in} > 50$, which we interpret as the failure of the fitting to converge, and so were discarded.  

 


\section{Conclusion}

Our primary finding is that the darkweb is a social network anomaly. Its light CORE and massive OUT components distinguishes it from other popular social networks. In fact, calling it a `web' is a connectivity misnomer; it is more accurate to view it as a set of dark silos -- a place of isolation, entirely distinct from the well connected world of the www and other social network. What causes the darkweb to be so isolated? We see two possible explanations:
\begin{itemize}
    \item \textbf{The technological explanation.} In the darkweb, sites go up and go down all the time.  Why bother linking if there's little chance that the destination will still exist?
    \item \textbf{The social explanation.}  People who create sites on the darkweb are cut from a different social cloth than those who create sites on the www (or at least when using the darkweb, these people behave differently)
\end{itemize}
\noindent
To test the technological explanation, we performed a second crawl collecting instances of darkweb sites linking to the www and compared the relative rates of outbound linking in Figure~\ref{fig:outdeg_as_func_of_indeg}(c). There are essentially equal rates of outbound linking to the www as well as the darkweb which tells us (i) the low outbound linking is not due to the impermanence of onion sites and (ii) if onion sites got drastically more stable, we would still see very low rates of linking. Taken together, these indicate the technological explanation is likely untrue. Thus, the social explanation is likely the cause of the darkweb's anomalous topology.

Rigorously testing the social hypothesis is however beyond the scope of this work. Although, in a sense we have taken a first step in this direction by generalizing preferential attachment which itself can be viewed as model of social behavior; it is a model of trust: highly linked nodes are perceived as `trustworthy' sources of information, and so receive a disproportionate number of links; the rich get richer. The isolated silos of the darkweb, however, indicate trust does \textit{not} play a role in the dynamics governing its evolution. Rather, one might say it indicates that \textit{distrust} does. The passive pages of the page and portal model (which recall, do not link to anybody through the dynamics, and are in that sense passive) were a crude way to incorporate this effect. But a more principled behavioral model (i.e. one consistent with known results from psychology) is needed, which were were unable to develop. We hope psychology-fluent researchers will take up this task in future work. 

Future work could also study the temporal aspects of the darkweb. Is the topology we have found stationary? For example, in the work most closely related to ours \cite{darknet}, it was found that the resilience of the studied `darknet' evolved over time (as discussed in the Data Collection section, our darkweb graph is much different to the darknet in \cite{darknet}).  It would be interesting to see if the resilience of our darkweb graph behaves like this too.

\section*{Acknowledgment}
The authors would like to thank all members of the MIT Senseable City Lab consortium.

%
%

\end{document}